# Low temperature growth and optical properties of α-Ga$_2$O$_3$ deposited on sapphire by plasma enhanced atomic layer deposition.

J.W. Roberts (1), P.R. Chalker (1), B. Ding (2), R. A. Oliver (2), J.T. Gibbon (3), L.A.H. Jones (3), V.R. Dhanak (3), L. J. Phillips (3), J. D. Major (3), F. C-P. Massabuau (2)

(1) University of Liverpool Centre for Materials and Structures
(2) University of Cambridge Department of Materials Science and Metallurgy
(3) University of Liverpool Stevenson Institute for Renewable Energy

Plasma enhanced atomic layer deposition was used to deposit thin films of Ga$_2$O$_3$ on to *c*-plane sapphire substrates using triethylgallium and O$_2$ plasma. The influence of substrate temperature and plasma processing parameters on the resultant crystallinity and optical properties of the Ga$_2$O$_3$ films were investigated. The deposition temperature was found to have a significant effect on the film crystallinity. At temperatures below 200°C amorphous Ga$_2$O$_3$ films were deposited. Between 250°C and 350°C the films became predominantly α-Ga$_2$O$_3$. Above 350°C the deposited films showed a mixture of α-Ga$_2$O$_3$ and ε-Ga$_2$O$_3$ phases. Plasma power and O$_2$ flow rate were observed to have less influence over the resultant phases present in the films. However, both parameters could be tuned to alter the strain of the film. Ultraviolet transmittance measurements on the Ga$_2$O$_3$ films showed that the bandgaps ranges from 5.0 eV to 5.2 eV with the largest bandgap of 5.2 eV occurring for the α-Ga$_2$O$_3$ phase deposited at 250°C.

## INTRODUCTION

Recent interest has been focused on gallium oxide (Ga$_2$O$_3$) because of its wide bandgap, which has been reported to be in the range from *ca.* 4.5 to 5.3 eV [1–6], making it a promising candidate for deep ultraviolet (UV) sensing applications [7,8]. Ga$_2$O$_3$ can exist as different phases (α, β, γ, ε, κ, δ), some of which are shared with In$_2$O$_3$ and Al$_2$O$_3$ [9–11], which have bandgaps of approximately 3.7 eV and 8.8 eV respectively [12,13]. Consequently, growth of In- or Al- doped Ga$_2$O$_3$ opens a viable route to bandgap tuning over a wide range of energies [14]. α- and β-Ga$_2$O$_3$ have also been identified as next generation candidates for the current carrying channels in power transistors due to their higher critical field strength compared to GaN and SiC [7].

To date, monoclinic β-Ga$_2$O$_3$ has been the most extensively studied phase, due to its stability compared to the other polymorphs and the ease with which it can be grown [10]. High quality crystals, polycrystalline films and nanostructures of β-Ga$_2$O$_3$, with bandgaps of *ca.* 4.9 eV [5], have been grown directly from melt formation [7], by chemical vapour deposition (CVD) [15,16], pulsed laser deposition (PLD) [3,17] and metal-organic chemical vapour deposition (MOCVD) [4,18,19], to name just a few examples.

α-Ga$_2$O$_3$, a metastable hexagonal phase of Ga$_2$O$_3$, has a wider bandgap than β-Ga$_2$O$_3$, around 5.1 eV - 5.3 eV [5,8] and is structurally analogous to α-Al$_2$O$_3$ (corundum)[11]. However, unlike the monoclinic β-phase, which is thermodynamically favourable, there are few processes which can reliably form α-Ga$_2$O$_3$, and all of them require temperatures above *ca.* 430°C [5] or high pressures of several GPa [20]. These requirements have limited the use of this polymorph in temperature and pressure sensitive processes and devices. Currently, the most prevalent method used for the growth of α-$Ga_2O_3$ is the "mist-CVD" technique that has been used to directly grow α-$Ga_2O_3$ between *ca.* 430°C and *ca.* 470°C [5,21,22]. Other synthesis methods include RF sputtering [23], precipitation [24], halide vapor phase epitaxy (HVPE) [25] and molecular beam epitaxy (MBE) [8]. Most $Ga_2O_3$ atomic layer deposition (ALD) papers using a variety of deposition temperatures, chemical precursors and substrates report deposition of amorphous material [26–32]. However, in a recent study we reported that α-$Ga_2O_3$ could be deposited by plasma enhanced ALD (PEALD) onto sapphire (α-$Al_2O_3$) substrates [33]. Here, we investigate the impact of several growth parameters (substrate temperature, plasma power and $O_2$ flow) on $Ga_2O_3$ thin films deposited by PEALD on sapphire.

**GROWTH METHODS**

Deposition of $Ga_2O_3$ films was carried out using an Oxford Instruments OpAL PEALD reactor with the baffle plate removed from above the chamber, giving a direct line-of-sight from the remote plasma to the substrates. The depositions utilised an inductively coupled plasma system located approximately 60 cm above the deposition chamber. Adduct grade triethylgallium (TEGa) from Epichem was used as the gallium source and dry $O_2$ from BOC was used as the oxygen source. Argon from BOC was used for chamber purges and as the precursor carrier gas. The TEGa source was maintained at 30°C, with line temperatures into the reactor chamber held at 80°C and 90°C. For the lowest temperature deposition (120°C substrate) the chamber walls were held at 125°C, while the chamber walls were set at 150°C for all other growths. For each growth run, several *c*-plane sapphire samples with a miscut of 0.25±0.10° towards ($11\bar{2}0$) were positioned centrally in the reaction chamber along with Si(100) pieces. 500 PEALD cycles were used for the growth of each film. Initial ALD growth parameters were taken from Shih *et al* [31] in which they found a saturative growth regime and linear growth per cycle using 0.1 s TEGa dose times and 5 s $O_2$ plasma exposure.

For each experimental set, the following conditions were kept constant between growths: 0.1 s TEGa dose, 5 s TEGa purge, 5 s $O_2$ plasma duration, 5 s $O_2$ plasma purge. 100 sccm Ar was used as a carrier gas during the TEGa dose and as the purge gas to remove unreacted precursors from the chamber during the purge steps. Table 1 lists experimental conditions that were changed between growths. The base pressure in the chamber (with no process gases flowing) was *ca.* 10 mTorr. During the deposition processes the chamber pressure varied between *ca.* 80 mTorr (during the plasma steps) and 160 mTorr (during the TEGa dose). Sample analysis was conducted on the as-grown films with no further annealing steps.

Table 1: Summary of the growth conditions for each of the experimental sets.

| Experimental set | $O_2$ plasma flow (sccm) | $O_2$ plasma power (W) | Substrate temperature (°C) |
|---|---|---|---|
| Temperature | 20 | 300 | 120, 150, 200, 250, 300, 350, 400, 450 |
| $O_2$ flow | 10, 20, 40, 60, 100 | 300 | 250 |
| $O_2$ power | 20 | 25, 50, 100, 200, 300 | 250 |

## ANALYSIS METHODS

After deposition, the ellipsometric parameters, psi ($\Psi$) and delta ($\Delta$), were measured over a range of 430-850 nm for each sample using a Horiba-Yvon MM-16 spectroscopic ellipsometer. Film thicknesses and refractive indices were computed using a $Ga_2O_3$ Cauchy model based on optical values reported by Rebien et al [34]. Whilst $\Psi$ and $\Delta$ measurements yield accurate values of refractive index (the error bar is smaller than the datapoints in Figures 1(A), 6(A) and 7(A)), the film thickness obtained using this approach are estimated to be within ~10% accuracy, owing to the model itself. Measurements were carried out on both c-plane sapphire and Si(100) substrates.

Characterisation by X-ray diffraction (XRD) was carried out using a PANalytical Empyrean diffractometer with a Cu K$\alpha_1$ X-ray source ($\lambda$ = 1.5405974 Å [35]), a hybrid monochromator, and either a two-bounce Ge crystal analyzer (for 2$\theta$-$\omega$ and $\omega$ scans) or a PIXcel detector (for reciprocal space maps (RSMs)). The strain state of the $\alpha$-$Ga_2O_3$ films was obtained by measuring the peak position from RSMs recorded around the symmetric 0006 and asymmetric $10\bar{1}10$ reflections.

The surface topography of the samples was investigated by atomic force microscopy (AFM), using a Bruker Dimension Icon Pro microscope operated in peak force tapping mode. A SCANASYST-AIR tip, with a nominal radius of 2 nm, was used.

Transmission electron microscopy (TEM) was used to observe the structure of the films in cross-section. The samples were prepared using standard sample preparation that entails mechanical polishing followed by $Ar^+$ ion milling at 5 kV and 0.1-1 kV. Annular dark field scanning TEM (ADF-STEM) was conducted in a Tecnai Osiris operated at 200 kV. Scanning electron diffraction (SED), which involves the acquisition of an electron diffraction pattern at each probe position [36] as a convergent electron probe is scanned across the sample, was performed using a Tecnai F20 operated at 200 kV and retrofitted with a NanoMegas Digistar system. This system enables the simultaneous scan and acquisition of diffraction patterns with an external optical charge coupled device imaging the phosphor viewing screen of the microscope. The SED data was inspected and analysed using a Python library for crystallographic electron microscopy [37,38]. A series of diffraction contrast images were formed by plotting the intensity within a selected subset of pixels in the diffraction pattern as a function of probe position to form so-called 'virtual dark-field' images.

The optical transmittance and reflectance of the films grown on sapphire were measured using a Shimadzu SolidSpec-3700 dual beam UV-Vis spectrophotometer (Shimadzu Corp., Kyoto, Japan) and the bandgap $E_g$ was extracted using the Tauc method.

## RESULTS AND DISCUSSION

### Effect of growth temperature

Figure 1(A) shows the variation in thickness of the $Ga_2O_3$ films grown on sapphire with increasing substrate temperature, as determined from fitting of the spectroscopic ellipsometry data. Average and standard error values across three samples used in each growth run are shown. At 120°C and 150°C, the $Ga_2O_3$ films thickness and refractive index are ca. 30 nm and ca. 1.87, respectively. XRD (Figure 1(B)) shows that these films are amorphous, with no visible peaks between 2$\theta$ values of 37° to 44° (apart from the 0006 reflection from the $\alpha$-$Al_2O_3$ substrate). At 200°C an increase in growth rate is observed, which coincides with the appearance of a partially strained $\alpha$-$Ga_2O_3$ 0006 peak in the XRD scan. Due to the broad assumptions used in the $Ga_2O_3$ Cauchy ellipsometry model, this could be attributed to either a real increase in growth rate due to the change in crystallinity (see XRD in Figure 1(B)), a deviation in the refractive index beyond that shown in Figure 1(A), or a change in the surface morphology which the model fails to account for. Between 200°C and 300°C the film thickness decreases to *ca.* 31 nm (corresponding to a growth rate of *ca.* 0.6 Å/cycle) and the α-Ga$_2$O$_3$ film approaches a more strain-free state (Figure 1(B) inset). We also note in Figure 1(B) the presence of fringes on the α-Ga$_2$O$_3$ 0006 peak with increasing growth temperature. Such fringes are often interpreted as improved film quality or greater uniformity of film thickness. However, as can be seen in the AFM results (Figure 2), the samples exhibit a similar roughness between 200 and 350°C, indicating a comparable film thickness uniformity between these samples. Therefore, the fringes in the XRD scan are probably more indicative of the crystal quality. At 350°C and above, XRD shows a decrease in the relative intensity of the α-Ga$_2$O$_3$ peak and the appearance of peaks at lower 2θ values, which may be assigned to one or several of the following reflections: ε-Ga$_2$O$_3$ 0004, ε-Ga$_2$O$_3$ $2\bar{1}11$ or β-Ga$_2$O$_3$ $31\bar{1}$ (the low intensity of the peaks does not allow us to discern the phase with certainty). The thickness of the films also appears to increase (Figure 1(A)), reaching a maximum growth rate of 0.84 Å/cycle at 450°C. However, considering the AFM (Figure 2(C)) and TEM (Figure 3(C)) data of the sample grown at 400°C, which shows a rough surface with the film thickness varying between 28 and 33 nm, it is possible that the ellipsometry model's assumptions break down for the samples deposited at 350°C and above. The XRD peaks seen in Figure 1(B) also correlate well with the refractive indices (at 632.8 nm) calculated from the ellipsometry data, with all crystalline films showing values between 1.9 and 2.0 [26,31,1,39,40] and the relaxed α-Ga$_2$O$_3$ films (250°C to 350°C) showing higher refractive indices between 1.95 and 2 in agreement with the literature [41].

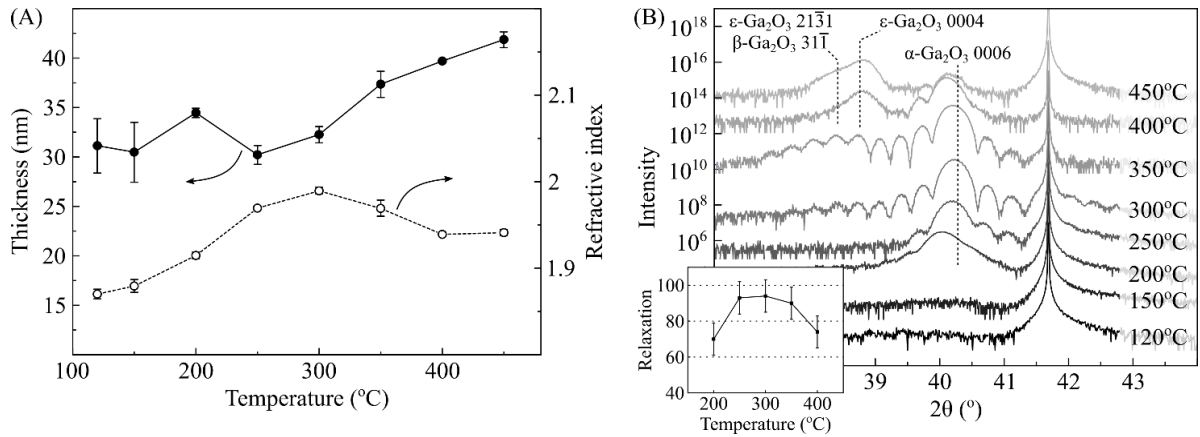

*Figure 1: (A) Ga$_2$O$_3$ film thicknesses and refractive indices (at 632 nm) against substrate temperature for growth on c-plane sapphire. (B) 2ϑ-ω scans recorded around the symmetric α-Al$_2$O$_3$ 0006 reflection showing the effect of growth temperature on the crystal structure of the Ga$_2$O$_3$ films. In inset, strain state of α-Ga$_2$O$_3$ with regard to the sapphire substrate at different growth temperatures.*

Figure 2 shows 500 nm x 500 nm AFM images of Ga$_2$O$_3$ films grown at 120°C, 250°C and 400°C, selected to represent each crystal phase observed, and a summary plot of the root mean squared (RMS) roughness for this sample set. At 150°C and below, the topography of the samples seems to indicate that the surface is amorphous and the RMS roughness values below 0.5 nm indicate a very uniform, conformal deposition. Between 200°C and 350°C, corresponding to the α phase, the films show a slight increase in roughness (between 0.5 nm and 0.7 nm) and the presence of grain-like structure with lateral size of around 10 nm (accurate measurement is impeded by the radius of the AFM tip).

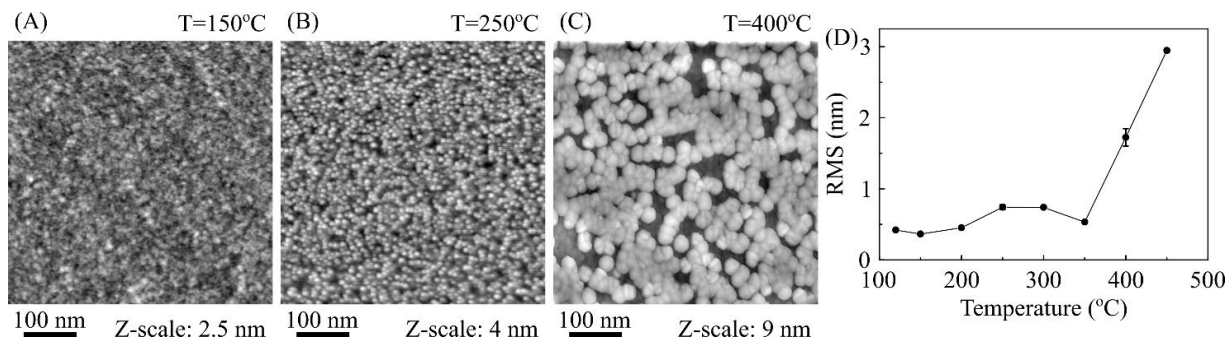

*Figure 2: 500 nm x 500 nm AFM images of the $Ga_2O_3$ films grown at (A) 150°C (amorphous $Ga_2O_3$), (B) grown at 250°C (α-$Ga_2O_3$), and (C) 400°C (mixed phases $Ga_2O_3$). (D) Plot of RMS roughness versus growth temperature.*

In our previous study of 130 nm thick $Ga_2O_3$ films, we observed by SED that the *ca.* 70 nm of $Ga_2O_3$ material adjacent to the substrate was dominantly α phase columns with a diameter of 2-23 nm[33]. Since the films shown here are approximately 30 nm thick, we expect that the grain-like structure observed here by AFM corresponds to the termination of the α phase columns. Above 350°C, the surface roughness increases by an order of magnitude up to a maximum of 3 nm at 450°C. The surface of the films at these temperatures features larger, less densely packed equiaxed features approximately 30 nm across and 5 nm high for the sample grown at 400°C, and 45 nm across and 8 nm high for the sample grown at 450°C. Given that these larger features appear in the samples where multiple phases cohabit (according to the XRD results (Figure 1(B))), it is reasonable to suggest that these large features correspond to grains of a different phase than the remainder of the film.

Figure 3 shows cross-sectional TEM images of $Ga_2O_3$ films grown on sapphire at 150°C, 250°C and 400°C, selected to represent each crystal phase observed. As can be seen in Figure 3(A), the film grown at 150 °C shows a uniform contrast and a smooth upper surface. The absence of features in this image support the XRD results, that is, that the $Ga_2O_3$ film is amorphous. The image also agrees with the AFM data which shows an average surface roughness of below 0.5 nm and the *ca.* 30 nm thickness determined by ellipsometry.

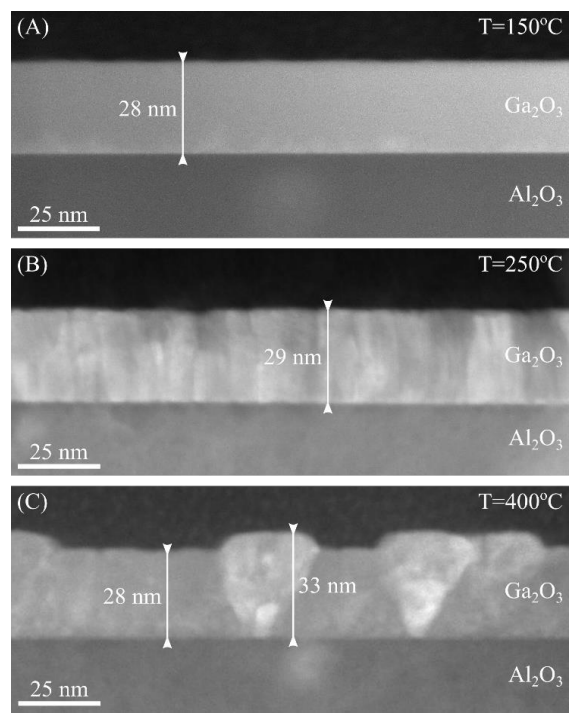

*Figure 3: Cross-sectional ADF-STEM images of the $Ga_2O_3$ films grown at (A) 150°C (amorphous $Ga_2O_3$), (B) grown at 250°C (α-$Ga_2O_3$), and (C) 400°C (mixed phases $Ga_2O_3$).*

The $Ga_2O_3$ film grown at 250°C is shown in Figure 3(B). Distinct columns of $Ga_2O_3$ can be seen running for the most part through the full thickness of the films. From our XRD results (Figure 1(B)), the film is dominantly α-$Ga_2O_3$. The structure of this film, that is, α phase columns, agrees well with our earlier report [33]. It can be seen that in comparison with Figure 3(A), there is a slight increase in surface roughness related to the grain boundaries where α-$Ga_2O_3$ columns touch. This is also in agreement with the AFM data (Figure 2). The film thickness (29 nm approximately) also agrees with the ellipsometry measurements (Figure 1(A)).

Figure 3(C) shows the Ga$_2$O$_3$ film grown at 400°C. The structure of this film is distinctly different from the first two images with large inverted triangular features visible. The space between these triangles is filled with material with a different contrast. Given that the whole film is expected to have the same average atomic number, contrast here arises from change of crystallinity in the film (crystal structure, orientation, *etc.*). The dimensions of the triangular features when they meet the free surface of the film agree very well with the dimensions of the equiaxed features we observed by AFM (Figure 2(C)). In light of the XRD and ADF-STEM results, we can safely say that these triangular features and the space between them relate to different phases of Ga$_2$O$_3$. Phase identification was obtained using SED, as shown in Figure 4(A). In this virtual dark field image, we can clearly attribute the inverted triangular features to α phase crystallites (a typical diffraction pattern recorded in such region is shown in Figure 4(B)) while the film between such features is mainly in the ε phase (a typical diffraction pattern recorded in such region is shown in Figure 4(C)). It should be noted that while the α-phase triangular features appear to be monocrystalline, the ε-phase material seem to be made from several small crystallites – which could explain why the corresponding XRD peak (around 38-39° in Figure 1(B)) is broad and weak. SED also confirms that the α-Ga$_2$O$_3$ features are grown along (0001) – which is the same orientation as the α-Al$_2$O$_3$ substrate. The ε-Ga$_2$O$_3$ crystallites are also dominantly oriented along (0001) although in places we observed signs of misorientation.

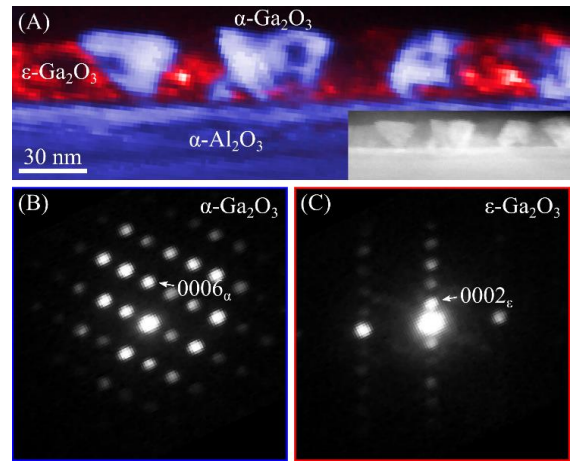

*Figure 4: (A) Virtual dark-field image obtained by SED of the Ga$_2$O$_3$ film grown at 400°C, seen along the $[11\bar{2}0]$ zone axis for α-Al$_2$O$_3$. In inset, corresponding bright field image. The blue colouring is obtained from the α phase – a typical diffraction pattern is shown in (B), and red colouring represents the ε phase – a typical diffraction pattern is shown in (C).*

Figure 5(A) shows the % transmittance against wavelength for the Ga$_2$O$_3$ films grown on sapphire. One data set for each morphology is shown for clarity. In the visible range, all films show similar transmittance at around 80%.

The fundamental electronic bandgap of β-Ga$_2$O$_3$ (located on the I-L line in reciprocal-space) is indirect with a magnitude of 4.84 eV, according to density functional theories simulations and experimental results obtained by other groups. However, several studies have also shown that the direct bandgap (located at Γ) is only 0.04 eV larger, with a bandgap of 4.88 eV [42–44]. The majority of papers available in which UV-Vis spectroscopy has been used for the study of Ga$_2$O$_3$ films use the direct bandgap assumption when calculating Tauc plots [1,4,26,28,30,32,45], *i.e*:

$$\alpha h\nu \propto \left(h\nu - E_g\right)^{1/2}$$

where $\alpha$ is the optical absorption coefficient, $h\nu$ is the photon energy and $E_g$ is the optical bandgap. Thus a plot of $(\alpha h\nu)^2$ against $h\nu$ can be used to extract $E_g$. Studies on the band structure of α-Ga$_2$O$_3$ are less readily available than those on β-Ga$_2$O$_3$, presumably due to the difficulty in synthesising high quality single crystals of the material. Choi *et al.* simulated doping of various materials into α-Ga$_2$O$_3$ and found that for the undoped oxide, the electronic bandgap was indirect with a value of 4.70 eV, whilst the nearest direct bandgap was 0.21 eV higher at 4.91 eV [14]. To conform with the existing literature, in the present work, we also use the direct bandgap assumption.

The inset in Figure 5(A) shows a Tauc plot for the same films assuming direct bandgaps [14,42–44]. Figure 5(B) shows the calculated optical bandgaps of the deposited films against deposition temperature. We observe an optical bandgap of *ca.* 5.05 eV for the amorphous Ga$_2$O$_3$ films (150-200 °C), while the α-Ga$_2$O$_3$ films (250-350 °C) exhibit a greater optical bandgap, nearer 5.15-5.20 eV. The bandgap of the mixed phase films (400-450 °C) is slightly lower than that of the purely α-Ga$_2$O$_3$ films, presumably due to the contribution of the ε phase. These bandgaps are in agreement with those found by other groups using a variety of different growth methods [5,19,21,40,46]. We observe that α-Ga$_2$O$_3$ exhibits the widest bandgap of all polymorphs, in agreement with the literature [19].

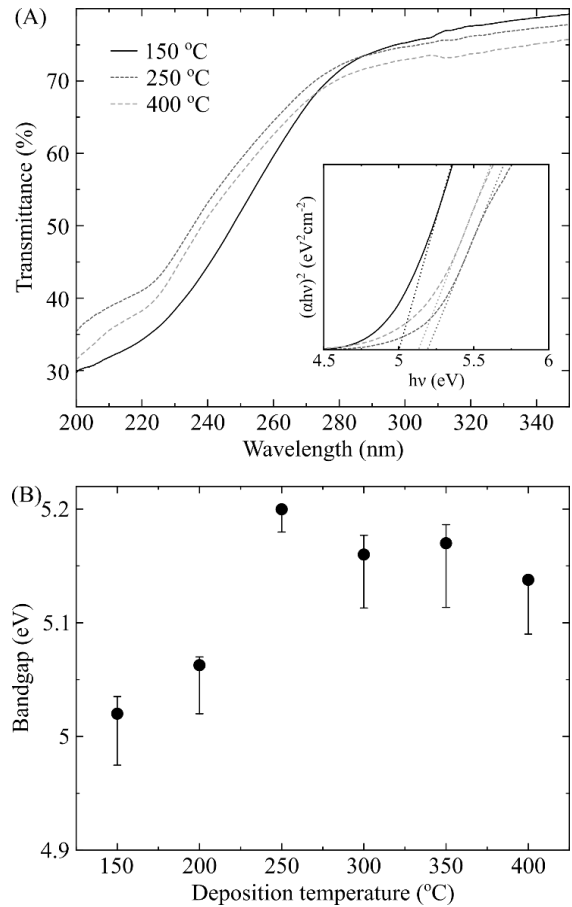

*Figure 5: (A) UV-vis transmittance for selected Ga$_2$O$_3$ films grown on sapphire (Inset – Tauc plot) and (B) Optical bandgaps for Ga$_2$O$_3$ films calculated from (A), assuming direct bandgap for each phase.*

*Effect of O$_2$ flow and plasma power*

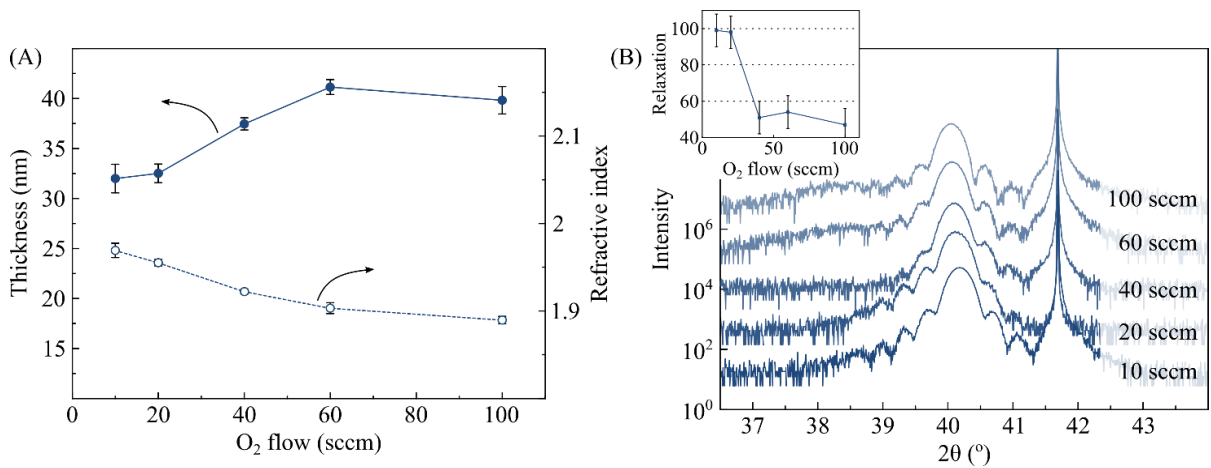

*Figure 6: (A) Spectroscopic ellipsometry model results add (B) XRD 2ϑ-ω scans of the samples grown under varying O$_2$ plasma gas flow at 300 W and 250°C substrate. In inset of (B), strain state of α-Ga$_2$O$_3$ with regard to the sapphire substrate at different O$_2$ flow.*

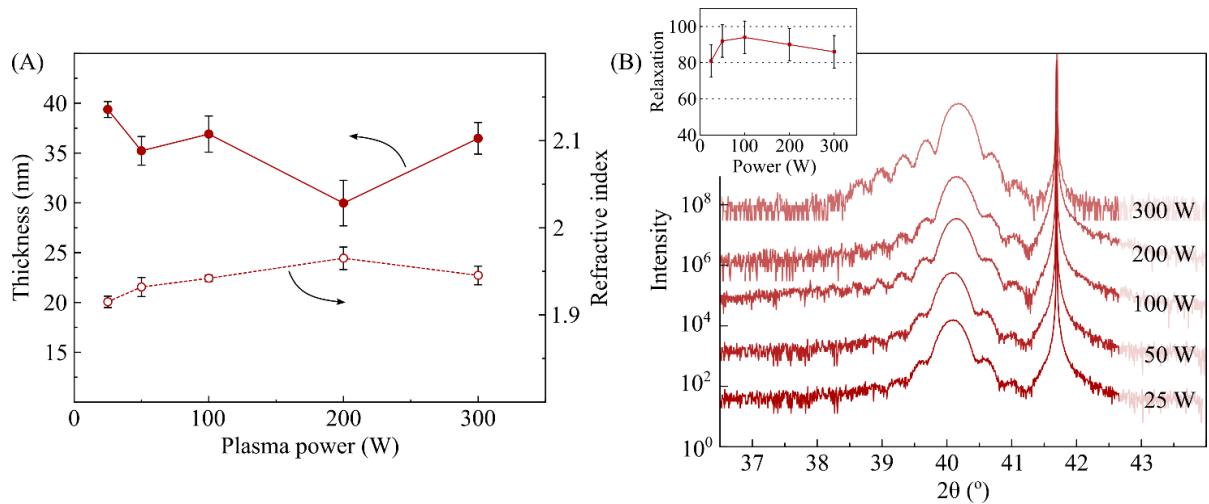

*Figure 7: (A) Spectroscopic ellipsometry model results add (B) XRD 2ϑ-ω scans of the samples grown under varying plasma power at 20 sccm $O_2$ flow and 250°C substrate. In inset of (B), strain state of α-$Ga_2O_3$ with regard to the sapphire substrate at different plasma power.*

Figures 6 and 7 show the spectroscopic ellipsometry and XRD results for varying $O_2$ flow and plasma power at a fixed growth temperature of 250°C for films grown on sapphire. The XRD results show that these two parameters have very limited impact on the crystal structure of the films – all films are in the α phase. The main noticeable difference was in terms of relaxation of the film (see insets Figures 6(B),7(B)). For the $O_2$ flow sample set, we observe that the lowest $O_2$ flows result in higher refractive indices and lower growth rates (Figure 6(A)) while also showing the highest strain relaxation close to 100% (Figure 6(B) inset). In comparison, films grown with higher $O_2$ flows show relaxation at around 50% – at the moment the mechanisms leading to strain relaxation in these films remain unclear. A higher refractive index and lower growth rate both imply that a denser film is being grown under the low $O_2$ flow conditions. AFM was also carried out on the films (not shown) but the change in these deposition parameters had no observable effect on the surface morphology or roughness – all the samples look similar to that presented in Figure 3(B).

## CONCLUSIONS

PEALD can be used to deposit α-$Ga_2O_3$ onto *c*-plane sapphire substrates at low temperatures and without the need for further annealing processes. A growth window for the α phase has been determined between 250°C and 350°C, with temperatures below the growth window resulting in amorphous $Ga_2O_3$ and temperatures above the growth window yielding mixed α and ε phases films. UV-vis transmittance measurements of α-$Ga_2O_3$ films on sapphire showed that the α-$Ga_2O_3$ film exhibits the highest optical bandgaps (up to 5.2 eV) amongst the samples studied. For α-$Ga_2O_3$ grown at 250°C, variation in the $O_2$ flow rate and $O_2$ plasma power growth parameters allow for further tuning of the strain relaxation of the α-$Ga_2O_3$ films with respect to the sapphire substrate.

## ACKNOWLEDGEMENTS

This project is funded by the Engineering and Physical Sciences Research Council (Grants No. EP/P00945X/1, EP/M010589/1, EP/N014057/1 and EP/K014471/1).